\documentclass[conference]{IEEEtran}

\usepackage{enumerate}
\usepackage{amssymb}
\usepackage[cmex10]{amsmath}
\usepackage{epsfig}
\usepackage{algpseudocode}
\usepackage{algorithm}
\usepackage{url}
\usepackage{color}
\usepackage{cite}
\usepackage{epstopdf}
\usepackage{amsthm}
\usepackage{colortbl}
\usepackage{bbm}
\usepackage{subfig}

 \definecolor{LightBlue}{rgb}{0.70,0.74,1}

\begin{document}

\title{Towards 6G Communications: Architecture, Challenges, and Future Directions}

\author{
 \IEEEauthorblockN{Purbita Mitra\IEEEauthorrefmark{1}, Rouprita Bhattacharjee\IEEEauthorrefmark{1}, Twinkle Chatterjee\IEEEauthorrefmark{1}, Soumalya De\IEEEauthorrefmark{1}, Raja Karmakar\IEEEauthorrefmark{1}, \\Arindam Ghosh\IEEEauthorrefmark{1}, Tinku Adhikari\IEEEauthorrefmark{1}}
 \IEEEauthorblockA{\IEEEauthorrefmark{1}Department of IT, Techno International New Town, Kolkata, INDIA 700156\\ 
 Emails: purbita.mitrahabra@gmail.com,roupritabhattacharjee@gmail.com,ctwinkle2812@gmail.com,\\soumalya384@gmail.com,rkarmakar.tict@gmail.com,happymailingtome@gmail.com,tinku.adhikari@gmail.com}
 }

\maketitle

\begin{abstract}

The cellular network standard is gradually stepping towards the 6th Generation (6G). %Though world of networking have not stepped into 5G fully the limitations of 5G technology is prominent which motivates the evolution of 6G. 
In 6G, the pioneering and exclusive features, such as creating connectivity even in space and under water, are attracting Governments, organizations and researchers to spend time, money, effort extensively in this area. In the direction of intelligent network management and distributed secured systems, Artificial Intelligence (AI) and blockchain are going to form the backbone of 6G, respectively. However, there is a need for the study of the 6g architecture and technology, such that researchers can identify the scopes of improvement in 6G. Therefore, in this survey, we discuss the primary requirements of 6G along with its overall architecture and technological aspects. We also highlight crucial challenges and future research directions in 6G networks, which can lead to the successful practical implementation of 6G, as per the objective of its introduction in next generation cellular networks.

\end{abstract}

\IEEEpeerreviewmaketitle

\section{Introduction}
\label{sec:intro}

%From basic wireless voice calling features of the 1st Generation (1G) to switching over to the voice over Long-Term Evolution (LTE) feature of the 4th Generation (4G), we have seen the launch of a new powerful generation every decade in the domain of mobile communications~\cite{david20186g}. 5G networks mark the starting of a true digital community and achieve a significant breakthrough in terms of data rates, latency, number of connected mobile devices, and mobility, compared to previous generations~\cite{soldani2015horizon,huang2019survey}.
 
Industry and academia are now working together in the direction of the 6th Generation (6G) cellular networks~\cite{chowdhury20206g,sun2020machine,jiang2021road}. Unlike the previous generations, 6G is expected to utilize a wide range of Artificial Intelligence (AI) services from the core network to terminal devices, leading to the concept of ``connected intelligence''~\cite{letaief2019roadmap,huang2019survey}, along with the optimization of 6G networks.
In the near future, we need connections underwater and in space. We need nanosecond response time such that the network is fast enough to download 100 hours of Netflix in a second. The network should be able to support 10 times more devices which are 100 times more reliable~\cite{jiang2021road}.
%In 5G, higher frequency waves, known as millimeter waves, can be used. However, they can be easily blocked like a tree. Thus, we have to set one big central tower with 4G and a tower in every town to get 5G, and as per T-Mobile, millimeter-wave 5G might never be scaled beyond the dense urban places. The 5G waves have to be compensated with sub-six waves and consideration of a mobile phone being able to switch between 5G, 4G and 3G should also be made so right now it seems like to deliver 
6G networks are expected to use sub-millimeter or terahertz waves, leading to an extremely high data rate~\cite{hong2021role,yan2020hybrid}. %If we cannot get the millimeter waves out of our cities how we are going to get it across countries. There are challenges of implementing 6G which needs an even shorter wave than 5G (1/10th of 5G waves). 
Samsung recently suggested about the incorporation of artificial intelligence by merging AI into 6G frameworks. It means the devices connected in the network can automatically talk to each other and smartly manage data between them. These modern technologies will help in constructing a stable communication source like 6G. 

When the 5th Generation (5G) cellular network was built, it had the idea to upgrade its capacity, so 6G could use the existing towers and frameworks as opposed to starting from the scratch. Another important consideration must be awareness of the damage we are causing to the planet and since we are trying to slow down the pollution level, 6G might be allowed to use as much energy as 5G would be using.
%We think designing a proper 6G network system, scheduled for 2030 but already started to be worked upon, would assist in data traffic management, increased data privacy, under-water network connections, etc. 
The rising in data traffic and ever increasing number of connected devices are the main driving forces behind the evolution of the next generation cellular network i.e., 6G~\cite{wikstrom2020challenges,jamil2020intelligent}. Therefore, we need an informative survey on the basic architecture and technologies of 6G network systems. The key features of the 6G network system are expected to be -- extreme low latency, ultra high data rates, high energy efficiency, enhanced security, and ubiquitous network coverage across the globe~\cite{jiang2021road}. Another prominent objective of the 6G network system is to satisfy the needs and requirements of Internet of Things (IoT) applications and uplift the widespread application and usage of the latest technological trends~\cite{huang2021accurate}. There has been a significant stress on the application of distributed ledger technology, such as blockchain, in the 6G network system as it is capable of providing privacy and security, which are the necessities of the digital world in the era of the Internet~\cite{kalbande20196g,nguyen2020privacy}. 

\textbf{Objective of this survey:} In this survey, we discuss the overall proposed architecture and technologies for 6G networks. %The new architecture of 6G targets at achieving higher network coverage than the current cellular technologies. 
We expect three tiers of network to the existing terrestrial network -- air tier network, space tier network, and underwater tier network. In the new architectural design of the 6G network, we will find extensive use of artificial intelligence and machine learning in designing unmanned aerial vehicles (UAVs), flying base stations, satellites, and appreciably more advanced forms of technologies. %It targets at making high speed networks available even to the remote areas world-wide, and the networks will also ensure the enhancement of security and trust in the network. 
We also present the challenges in the design and deployment of 6g networks. Based on the overview of architecture and technologies of 6G, we also discuss future research scopes in 6G, which can help practically implement the proposed 6G network, leading to the intelligent network management with more security and data rates compared to the existing cellular networks. Consequently, 6G will help in merging the economically challenged population with the technologically advanced world, that leads to the opening of a wide range of benefits and possibilities to ease the way of living tomorrow.

\textbf{Organization of this survey:} This survey is organized as follows. Section~\ref{sec:architecture} discusses the overall architecture of 6G. Section~\ref{sec:technology} describes the technologies in 6G networks. The research challenges of 6G are presented in Section~\ref{sec:challenge}, and Section~\ref{sec:futurescope} highlights the future research scopes in 6G networks. Finally,~\ref{sec:conclusion} concludes the survey.

\section{6G Architecture}
\label{sec:architecture}

In this section, we present an overview of 6G architecture. 

\subsection{Vision of 6G Architecture}

\begin{figure}[!t]
	\centering
	\includegraphics[width=0.8\linewidth]{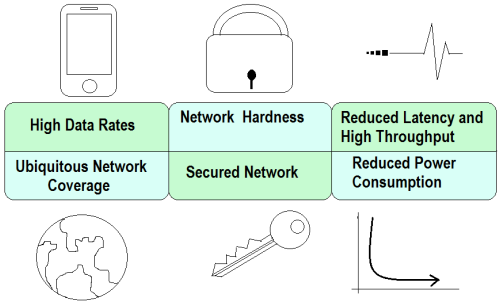}
	\caption{Primary requirements of 6G networks}
	\label{fig:arch1}
     \vspace{-5mm}
\end{figure}

6G mobile networks target ubiquitous intelligence, computing power, and high-speed wireless connectivity throughout air, space, and sea. The vision to achieve this objective is integrating underwater communications and satellite communication networks to provide network coverage throughout the globe~\cite{huang2019survey,zhang20196g}. 
There is a need for a super speedy service in 6G mobile networks with data speed close to about $1000$ Mbps~\cite{kalbande20196g,yu2020cybertwin}. %Although the requirements of 6G network are still being investigated, 
Some of the requirements of 6G can be listed as -- holographic communication, ultra high broadband, multi-sense transmission, ultra-high throughput, reliability, low latency, etc.~\cite{huang2019survey,wang20206g}. Fig.~\ref{fig:arch1} shows the primary requirements of 6G networks~\cite{yaacoub2020key}.

Probable solutions can be using smaller cell sizes and higher frequency bands. However, smaller sizes of cells will lead to more power consumption and high operational costs, and high frequency bands can suffer path loss. Therefore, we have to put a limit to decrease the size of cells and increase the frequency bands. Fully-decoupled radio access network (FD-RAN) architecture has been proposed, where the network functionalities are fully decoupled and will be deployed by every independent network entity. Implementation of multi-point coordination and centralized resource management can obtain an elastic resource cooperation in a fully decoupled RAN. The control base station of FD-RAN co-ordinates with decoupled uplink and decoupled downlink in a centralized way, which has a similar architecture as cloud-RAN~\cite{yu2020cybertwin}. 6G will use a very wide variety of computing, networking, communications, and many sensing technologies to offer smart applications. The key enablers are AI, edge-intelligence, blockchain, homomorphic encryption, network slicing, and integrated network for space, sea, and ground~\cite{yan2020interference,khan20206g,aggarwal2020blockchain}.

\subsection{4-Tier Ubiquitous Coverage of 6G Networks}

There are 4-tiers in the ubiquitous coverage of 6G networks, as follows.

\subsubsection{Space Network Tier}
 
The space network tier will support wireless space Internet services using densely deployed low earth orbit (LEO), medium earth orbit MEO), and geostationary earth orbit (GEO) satellites~\cite{yan2020interference}. The geostationary satellites are at an altitude of 35,786km which leads to signal transmission delay as the nodes are at huge distances from each other. This raises a need for non-geostationary satellite orbits (non-GSO) which is expected to provide high-bit rates and low latency Internet connectivity throughout the globe~\cite{fang20205g}. Since the deployment of NGSO satellites will still take a lot of time, LEO satellites can be opted as a possible solution. LEO satellite network with radio frequency and laser co-routing mechanism will provide lower latency communications, where communication distances are above 3000km~\cite{huang2019survey,zhou2019design}. 

\subsubsection{Air Network Tier} 

The air network can be broadly divided into two categories- Low Altitude Platform (LAP) and High Altitude Platform (HAP). The LAP operates within an altitude of a few kilometres and the HAP generally operates
in the stratosphere. The HAP has longer endurance and wider coverage capabilities but the HAP overlaps with LEO
satellite networks~\cite{huang2019survey,zeng2016wireless}. On the other hand, LAP uses low frequency, Millimeter-wave (mmWave) bands, and microwaves for providing flexible and reliable connectivity by dense employment of flying base stations (BSs) and UAVs. The UAVs can act as relay nodes for long distance communications, and thus promote the integration of space and terrestrial networks. But the most interesting and helpful characteristic of UAVs is that it can facilitate mobile communications in areas where adequate infrastructures are absent~\cite{gupta2015survey}. The UAVs have high speed mobility, and therefore, more advanced mobile network protocols are required for UAVs. The adaptive hybrid communication has potential and surpasses the existing protocols~\cite{zheng2018adaptive}. A fully integrated 6G network can be expected to have multilayer architectures including UAV based LAPs, LEO satellites, GEO satellites, and heterogeneous terrestrial networks~\cite{huang2019survey,zhang20196g,aggarwal2020blockchain}.
 
\subsubsection{Terrestrial Network}

Terrestrial networks will be a highly dense heterogeneous network requiring the deployment of X-haul of high capacity as a large number of small BSs need to be deployed to prevent the path loss. Terrestrial networks will support microwave, mmWave, low frequency, and tetra hertz band~\cite{zhang20196g}. 

\subsubsection{Under-Sea Network} 

The under-sea network targets on providing Internet facility under the seas and oceans. But it still remains a very controversial topic whether or not it would be a part of 6G networks. This network system involves optical and acoustic communication, and radio frequency. But the difficulty lies due to the unpredictability of the underwater environment, challenges that are to be faced, and the risk factors as water exhibits different propagation properties which are different from that of aerial or terrestrial terrains. Therefore, a lot of issues are yet to be resolved for under water 6g networks~\cite{huang2019survey,zhang20196g}.

\subsection{New Network Protocols}

%The internet protocol has many challenges and it is not suitable for applications in further network architecture developments with futuristic approach. 
The current internet protocol architecture is not applicable for 6G due its limitations in terms of latency and throughput. Recently, some new protocols like Quick UDP Internet Connections (QUIC) have been built that have eased these issues to a certain extent. But, unfortunately these protocols make the Internet more complicated and cannot serve as a complete remedy~\cite{zhang2016smart}.
%Therefore there is still lies the need of  a venture  a need to come up with a better and reliable solutions with developed and advanced protocols to effectively provide the required services~\cite{zhang2016smart}.

\subsection{Large Dimensional Network Architecture}

The multi-connectivity technique can help in connecting multiple access network tiers to achieve enhanced coverage~\cite{nawaz2021non}. However, this kind of connectivity arrangement between aerial/space, and terrestrial network tier is non-ideal, and therefore lead to a large latency which in turn will affect the information transfer. Also, the high mobility of BSs makes it difficult to get the channel state information for the connectivity management. The BSs also may generate errors in channel estimation. Another problem that can arise is the interference among the different tiers. The possible solution to tackle the interference of tiers can be done by setting a schedule system where information will be shared to upgrade user scheduling. The authors in~\cite{zhang20196g} discuss the need for high-performance algorithms for the scheduling.

\section{Technologies in 6G}
\label{sec:technology}

\begin{figure}[!t]
	\centering
	\includegraphics[width=\linewidth]{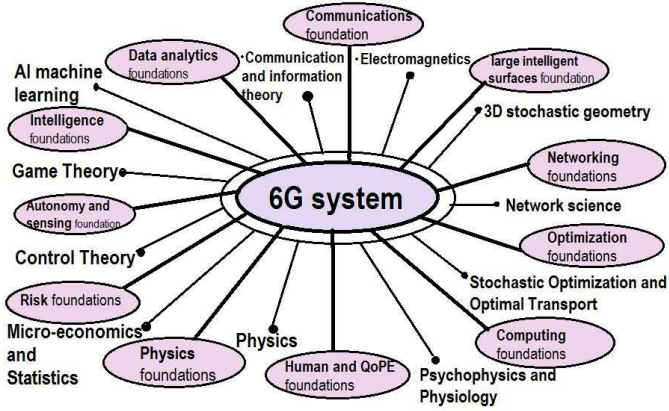}
	\caption{Different technological aspects of 6G}
	\label{fig:arch2}
     \vspace{-5mm}
\end{figure}

Spectrum communication technique for 6G mainly uses the THz band which is basically a spectrum band (0.1 THz to 10 THz) having many unique characteristics for communications. These waves have a narrow beam that secures the
communication, and the data rate is close to 100 Gbps. The spectrum communication is also suitable for broad applications such as high-speed wireless and space communications~\cite{huang2019survey}. %The global research made some breakthrough for some challenges which occurs in high level THz wireless communication systemssystem. 
A wireless transceiver chipset (data rate of 80 Gbps) is designed, which solves the problem of low antenna-gain~\cite{huang2019survey}. In 6G, the high frequency connection reduces many problems in many areas like cross-talks, reflection, etc.~\cite{huang2019survey}. Fig.~\ref{fig:arch2} shows different technological aspects of 6G networks.

\begin{figure}[!t]
	\centering
	\includegraphics[width=\linewidth]{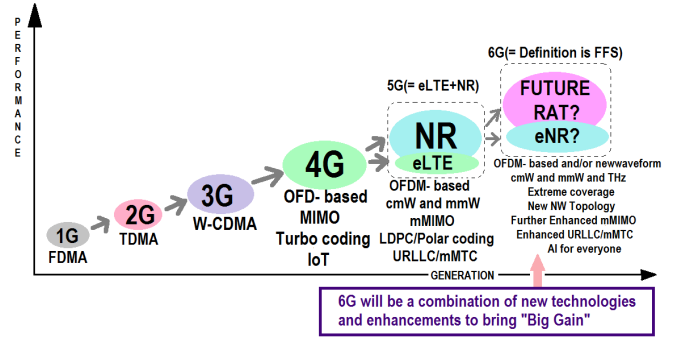}
	\caption{Performance enhancements with respect to cellular technology generations}
	\label{fig:tech1}
     \vspace{-5mm}
\end{figure}

The usage of a robust beamforming and scanning algorithm needs a significant research before use it properly in 6G. In addition, other parameters need to be investigated, which include low-complexity, low-power hardware circuits, channel and noise modeling, energy efficient modulation schemes, low density channel coding, ultra massive multiple-input and multiple-output (MIMO) systems, and powerful synchronization schemes~\cite{huang2019survey}. The visible light communication using optical wireless communication is one of the most preferable technologies in the 6G architecture~\cite{katz2020opportunities}. In 6G, the use of light-emitting diodes (LEDs) can help us in the high-speed data communication. %In this technology, the sharing of information in the network is more smoother and more preferrable than the and. free and unlicensed nature~\cite{huang2019survey}.
Fig.~\ref{fig:tech1} illustrates the technological enhancements of cellular networks, that lead to the era of 6G.

\subsection{New Communication Paradigm}

In molecular communication (MC), biochemical signals (usually small particles) create a gaseous medium to make the communication. It is usually compared with radio communications. MC signals are mainly biocompatible as they consumes very small energy particles. It mainly wants to interface with mobile networks and the Internet, which are more challengeable~\cite{huang2019survey}. Quantum communication (QC) is one of the most secure forms of communications. We encode the information in the quantum state by using photons or quantum particles. %The QC can improve data security and after several years of exploration, there will be many branches of QC, such as quantum teleportation, quantum secret, etc. 
The QC has a huge potential in long-distance communications~\cite{huang2019survey}. Therefore, both MC and QC can be effective in 6G~\cite{nawaz2019quantum}.

%Fig. Architecture of different privacy systems

\subsection{Homomorphic Encryption}

In homomorphic encryption, people perform cryptographic operations on some specific algebraic operation to get a truthful result.  
We want to combine machine learning (ML) and homomorphic encryption to perform tasks on encrypted data and get the result in plaintext~\cite{sun2020machine}. %The more common algo are cryptographic neural network and the researchers adjust the different algo into the application of ML.In this generation deep learning has huge number of fields to appeal to create its work, 
If we apply deep learning for wireless communications in 6G, it can give us a better result~\cite{sun2020machine}. We will use ML algorithm for learning and prediction whenever we are protecting data security in a specific biological atmosphere. We will also use homomorphic encryption for maintaining the privacy of the data along with computing time costs. 

\subsection{Free-Space Optical Communication}

Free-space optical communication (FSO) is a visible light communication to transform data, and it has very high level data rates, and thus it is tough to implement it in rural areas~\cite{yaacoub2020key}. FSO mainly provides the backhaul connectivity, and is also used for fronthaul access in rural areas~\cite{yaacoub2020key}. In FSO, there are mainly two challenges -- control of the Doppler effect for satellites and varying delays of user equipments (UEs) in the area of satellite~\cite{yaacoub2020key}.

\subsection{Scope of Artificial Intelligence}

Apart from low latency and high throughput, an important point of difference between previous networks and the new 6G
networks is intelligence. Therefore, there is a promising scope of introducing AI in the architectural framework of 6G autonomous networks. AI can learn, predict and then make rational decisions based on training. %This gives the ability to replace and transform fields for enhancing the performance of wireless networks. 
Hence, AI provides a contemporary solution for the technology enhancement and superior network performance. The features, such as network function virtualization (NFV), software-defined networking (SDN), and network slicing (NS), will continue to be parts of 6G networks as cloudization, slicing, softwarization, virtualization, etc. 

The NFV decouples the software and hardware sections to waive off the dependency of software on hardware and NS creates parallel network functions to achieve network deployment and management on demand. The function of SDN is to decouple the control function to achieve programmable configuration and network management. A junction of SDN, network slicing, NFV, and AI can bring about zero-touch and dynamic network composition. Self aggregation of different radio technologies can be achieved with the help of AI for liquidized networks and satisfactory demands of constantly changing applications and services. The authors in~\cite{zhang20196g} have clearly indicated the importance of AI in the architecture of the new 6g network as it is capable of real-time monitoring. AI will also escalate the network utility and improve the quality of experience. In order to promote the inclusion of AI in the 6G network design, machine learning must be the centre of attraction~\cite{zhang20196g}. 

\subsection{AI-Enabled Deployment and Energy Efficiency}

To distribute the enhanced quality of service (QS) for smartphone users, the 3D positioning of aerial BSs is used to assist ground stations. The energy-efficiency behaviour of 6G is an important study. In the direction of the 3D positioning and energy-efficient deployment, AI can be used to make a learning-based model. Specifically, the researchers use actor-critic reinforcement learning for BS users~\cite{elsayed2019ai}.

\subsection{Big Data Analytics for 6G}

6G can be plugged with the use of descriptive, diagnostic, predictive, and prescriptive analytics in the big data. The use of AI along with big data can lead to an intelligent 6G model handling a huge volume of data processing~\cite{letaief2019roadmap}.

\subsection{Brazil 6G Project} 

In Brazil, the 6G project started in 3 phases, as given in the following.~\cite{brito2020brazil}.
\begin{itemize}
 \item \textit{The art in 6G networks:} Here we find the architecture, vision, and application of the 6G research. 
 \item \textit{The scenario of Brazil for 6G:} This phase finds the applications which are suitable for
Brazilian environment. 
 \item \textit{The trends of technology:} This application finds the main required technology for 6G.
\end{itemize}

\subsection{Intra-Inter Chips Length Scale}

The optic wireless communication mainly depends on inter-intra-chip length scale, the ability of retransmission totally depends on the structure of the ability of the received signal~\cite{alves2020plasmonic}. This retransmission technique can be widely used to improve the performance of 6G.

\subsection{Combination of Cloud/Edge/Terminal Computing} 

The cloud computing has some advantages like it offers users for using the remote infrastructure. For this purpose, the mobile application is released with the minimal management efforts~\cite{zhu2020exploring}. The edge computing is basically a service based on the location to the edge of smartphone networks and wants to avoid the network traffic~\cite{zhu2020exploring}. There is a centralized server which gradually increases the communication between user-friendly devices. But we can further recognize more possibilities using the aforementioned computing mechanisms ~\cite{zhu2020exploring}.

\subsection{Uses of Blockchain in 6G}

For the spectrum sharing, orchestration, and decentralized computation, the application of blockchain is an important step. By the application of blockchain~\cite{hewa2020role,maksymyuk2020blockchain} with AI, the relation between users and operators can be controlled in a smart manner. The design of the main centralized access control system and the mechanism of radio clouding are integrated with blockchain~\cite{hewa2020role}. Big data computing can create lots of hazards in the environment, but using the application of blockchain, we can prevent the hazards~\cite{hewa2020role}.

\section{Challenges in 6G}
\label{sec:challenge}

Developers and researchers must think beyond 5G now as the adaptation of mankind to 5G will lead to new demanding challenges. Hence, to satisfy human needs, a novel network with a fresh vision has to be introduced. There is an urgent need for the development beyond 5G because there is a need of an intelligent network management with very high data rates. In order to be able to suffice the human needs of the next era and emerging challenges, a new network with new vision has to be introduced~\cite{wikstrom2020challenges}. One of the key objectives of the 6G network is to reach every corner of the world, irrespective of terrain, other environmental factors and social strata of the people. 
The population of unconnected people is mostly concentrated over rural regions with low per capita income and low literacy rates. 
Next, we discuss different challenging issues in 6G.

\subsection{Internet Connectivity and Telecommunication Infrastructure}
There is a sharp disparity in terms of Internet connectivity in urban and rural areas because the rural areas lack the availability of required infrastructure (e.g. water, electricity, transportation, etc.)~\cite{yaacoub2020key}. The difficult terrains that can be encountered in rural areas make the setting up of required infrastructure more difficult due to out-of-reach terrains in rural areas and maintenance of various equipments. Therefore, the basic challenges in 6G, that may be encountered for setting up the telecommunication infrastructure, can be summed up as: 
\begin{enumerate}
 \item Insufficient business due to sparse population,
 \item Increased capital expenditure (CAPEX) due to deployment of base stations. 
 \item Little or no electricity service leading to high operational expenditure (OPEX) for setting up diesel
generators in base stations.
 \item Lesser skilled personnel leading to difficulty in maintenance~\cite{yaacoub2020key}. 
\end{enumerate}

\subsection{Security and Varieties in Connected Devices}
Moreover, there is a rapid increase in the number of varieties of connected devices due to the inclusion of IoT, low latency, and higher throughput. Therefore, a significant improvement in cellular networks is required in order to cope with IoT, and meet the ever increasing speed and other features such as network densification, societal automation, public safety, reliability, etc.~\cite{wikstrom2020challenges,huang2021accurate}. Recently, blockchain can be an efficient secure information exchange technology because of its features like decentralized architecture and elimination of central third parties, transparency with anonymity, less processing delay, etc. Therefore, the appropriate application of blockchain in 6G networks can be a great challenge. 

\subsection{Intelligent Resource Management and Extreme Mobility}
The intelligent resource management of networks can be challenging considering the expected enormous connectivity demands in the telecommunication ecosystems in future mobile communications. In 6G, the operations, such as decentralized computation, spectrum sharing, and orchestration, need to be compatible with a huge infrastructure~\cite{zhang2019edge}. There will be an extremely high rise in the number of devices that will be operated in future and will be connected to the Internet, which raises the need of extremely high speed network infrastructure to match the demands. This makes the design and execution of 6G network architecture very difficult because of unprecedented traffic demand. Also, with the rising number of devices, there will be a rise in the various application usages and real-time communications are expected to emerge as a very urgent top-priority need in 6G, leading to life-critical communications~\cite{berardinelli20206g}. The aforementioned challenges can be diminished only with close-to-zero delays and a high accuracy. The base stations should be able to handle this exchange of data on a large scale. In addition, the synchronization of vehicular networks, electricity supplies, and many other services are needed to serve the enormous rise in the connected devices and application usages. All these make the configuration of world-wide 6g networks a challenging task. 

\subsection{Overall Challenges}
\begin{figure}[!t]
	\centering
	\includegraphics[width=0.9\linewidth]{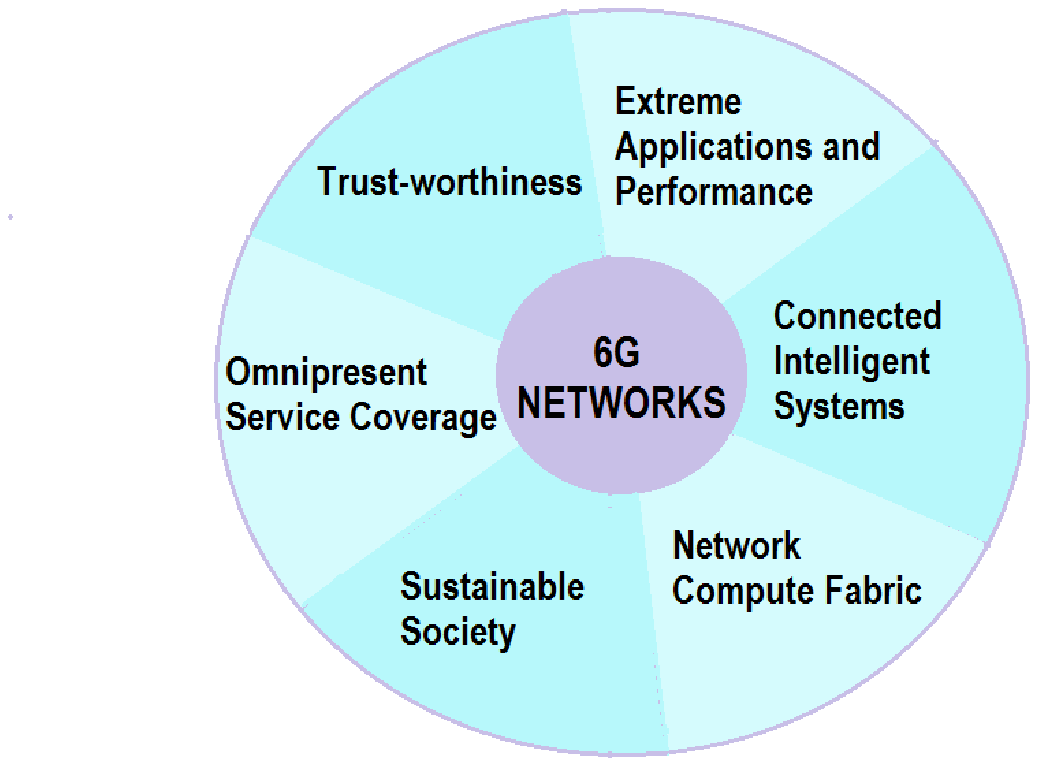}
	\caption{Challenges of 6G in different directions}
	\label{fig:chlg}
     \vspace{-5mm}
\end{figure}

Fig.~\ref{fig:chlg} shows the challenges of 6G in various directions.
Overall, few important queries that arise as challenges are as follows.
\begin{itemize}
 \item \textbf{Trustworthiness:} Although there are many security features in 5G networks, there are requirements of a better security, reliability, high predictability, and high resilience of delivered functionality, which will be highly demanded from the user-end in 6G. Network should be able to revert back any activity of jamming, tampering or intrusion. 
 \item \textbf{Omnipresent service coverage:} 6G networks are expected to provide service on land, air, water, and space, with strong and secured network connections. Along with these features, 6G networks are expected to meet and tackle the ever increasing demand of high speed services. All these expectations need 6G networks to be robust and versatile in providing network services. 
 \item \textbf{Performance and extreme applications:} 6G is expected to process the vast amount of information, and simultaneously serve the high speed network with the gradually increasing number of connected devices. The low latency, full sensory communication, reduced power consumption, and extended range of network availability are some of the key features of 6G networks. Moreover, 6G will hopefully be used in the fields of agriculture, e-health, law enforcement, governance, transportation, entertainment, and many daily-life applications. 
 \item \textbf{Connected intelligent systems:} Artificial intelligence and machine learning tools will become a significant part of connected integrated systems. Presently, they might be used to replace conventional solutions like brute force algorithms or heuristic approaches. However, as the days pass by the development, AI and machine learning applications in zero-touch operations, real-time analysis, multi-connectivity setup, mobility, etc. are going to take up larger spaces in the connected integrated 6G network systems.
 \item \textbf{High quality of experience:} A huge number of use cases poses varying requirements in 6G, which include extreme mobility, cost-effectiveness, high battery lifespan, and ultra-low latencies. To incorporate all the expected features of the intelligent management of the high speed mobility, 6G architecture has to be flexible to support huge number of online learning-based processing instances, variety of deployments, and consequently the quality of experience (QoE) will be significantly increased in 6G. 
 \item \textbf{Sustainability:} 6G is expected to be a big contributing factor to fulfil sustainable development goals and socioeconomics transformations. A significant reduction in carbon emissions, energy-saving wireless terminals, and helping to streamline other industries are some of the targets of 6G networks~\cite{wikstrom2020challenges}.
\end{itemize}

\section{Future Research Directions in 6G}
\label{sec:futurescope}

In this section, we present future research scopes in 6G.

\subsection{Applications of Machine Learning Algorithms}

Since machine learning is an important domain of 6G, we must explore different machine learning algorithms to examine the impacts of them on different performance factors of 6G. Generalising the machine learning process will not provide satisfactory results for frequently added data in an application. We must keep preferable federated learning on the top priority list in machine learning applications, however they also suffer from fairness issues~\cite{liu2020federated,khan20206g}. The transfer learning can be an efficient-approach for making the 6G system adaptable with the wireless environment~\cite{wang2021transfer}.

\subsection{Scalable and Reliable Blockchain-Enabled 6G}

For secure storage in various smart services like immutable ledgers and transactions in a distributed way, we generally use blockchain in 6G~\cite{maksymyuk2020blockchain}. Blockchain is also used in various smart services in health care, smart supply chain management, etc., where the 6G technology will have extremely low latency and low energy consumption along with maintaining the security perspective. We also want to enhance the scalability and reliability of the 6G system. Blockchain can be an efficient tool to address all the aforementioned features along with preserving privacy concerns~\cite{khan20206g,nguyen2020privacy}.

\subsection{Meta-Learning-Enabled 6G}

We want to experiment with specific machine learning techniques or algorithms for 6G applications. In this case, meta learning provides us machine learning models which learn the metadata of a machine learning-based experiment. We want to merge general machine learning with meta-learning, that can lead to a smarter technology for future 6G applications and evolution of wireless mobile technology~\cite{khan20206g}.

\subsection{Cloud-based Architecture and Technology}

Basically, 6G is an edge-centric and data-flow based technology. The chain of network functions and services is dynamically based on the optimal balance between the consumed and available resources in the cloud network. The system should be based on the combination of different machine learning approaches in the cloud environment. We will also focus on the privacy and security of the dataset used in the technology~\cite{khan20206g}.

\subsection{Terahertz Frequencies to Boost Data Rate}

We want to enhance the data rate in 6G applications, and the high data rate is one of the key requirements in 6G. For this purpose, we will use the frequency band above $52.6$ GHz with ultra-high rates of 100 Gbps and more~\cite{yan2020hybrid}. We will use the THz band to improve spectrum efficiency and mitigate free space loss, molecular absorption, etc.~\cite{ziegler20206g,yan2020hybrid}. We also need to control the energy-efficiency field to make it more suitable since integrating large numbers of antennas can increase the signal-to-noise ratio (SNR). By applying the multiple-input and multiple-output (MIMO) configuration, we can enable the cost-efficient delivery with the high data rate in a large area of 6G networks~\cite{chen2020wireless}. We also should take care of the electromagnetic field and bio-aware beam streaming field as well~\cite{ziegler20206g}.

\subsection{AI-based Edge Computing}

The increased number of data processing cannot be handled by the mobile computing which is still used in 5G. Therefore, the mobile edge computing is gaining a significant momentum, as it can distribute the network into a cloud computing architecture. We need to reduce the latency in 6G networks, where we can access the radio networks easily~\cite{tomkos2020toward}. We are planning to use more AI in the final devices or in the whole process of 6G
technology. We also need to focus on offline computing functions by using machine learning techniques~\cite{tomkos2020toward}, such that the resource allocation and offloading in the cloud and edge can be performed in an intelligent way~\cite{jamil2020intelligent}. We can use a central controller which divides the applications/dataset in the form of mini-batches and helps allocate in the multiple processing devices~\cite{tomkos2020toward} in the edge or cloud. In real-time communications, the cloudification (or edge clouding) is an expected technical goal of 6G connections. The use of AI and reusable data influence may help us reach the new expected economy scale and many other new scopes. We need to build an online learning-based connection between different components of 6G frameworks~\cite{ahokangas2020antecedents}. 
In agriculture and industries, 6G can intelligently handle the production generation, and impose secured and very fast connections between different modules.

\subsection{Use of Localization-based on an Intelligent System (LIS)}

LIS-aided mmWave systems can be used as positioning in 6G design. Here, we want to find the optimal location, by which we can lace the reflection and metasurfaces. This task is quite challenging, its a inverse problem of channel modeling. On the other hand, an efficient-deployment of materials in LIS needs to be explored~\cite{alghamdi2020intelligent}.

\section{Conclusion}
\label{sec:conclusion}

In this survey, we discuss the architecture and technology of 6G, where new features are introduced, focusing on the enhancement of the speed compared to the previous cellular standards.  
By using 6G, we are trying to reach everywhere like underwater, space, etc., where normally we cannot even imagine to reach. Based on machine learning and blockchain, 6G is expected to have an intelligent network management with a distributed nature of handling privacy and security. We also highlight the design challenges and future research directions in 6G networks, which can help identify the scopes of improvements in 6G systems, such that the successful deployment of 6G can be possible as per the objective of the invention of 6G.

\bibliographystyle{IEEEtran}
\bibliography{reference}

\end{document}